\documentstyle[twocolumn,aps,prl,epsf]{revtex}
\begin{document}
\title{Piezomagnetism and Stress Induced Paramagnetic Meissner Effect in 
Mechanically Loaded High-$T_c$ Granular Superconductors}
\author{Sergei A. Sergeenkov\cite{byline}}
\address{Theoretische Physik, ETH--H\"onggerberg, 8093 Z\"urich, Switzerland}
\address{\em (\today)}
\address{
\centering{
\begin{minipage}{16cm}
Two novel phenomena in a weakly coupled granular superconductor under an applied
stress are predicted which are based on recently suggested piezophase effect 
(a macroscopic quantum analog of the piezoelectric effect) in 
mechanically loaded grain boundary Josephson junctions. Namely, we consider the
existence of stress induced paramagnetic moment in zero applied magnetic field 
(piezomagnetism) and its influence on a low-field magnetization (leading to 
a mechanically induced paramagnetic Meissner effect). The conditions under which 
these two effects can be experimentally measured in high-$T_c$ granular superconductors
are discussed.
\end{minipage}}}
\maketitle
\narrowtext

Despite the fact that granular superconductors have been actively studied (both 
experimentally and theoretically) for decades, they continue contributing to the 
variety of intriguing and peculiar phenomena (both fundamental and important for
potential applications) providing at the same time a useful tool for testing new 
theoretical concepts~\cite{1}. To give just a few recent examples, it is sufficient 
to mention paramagnetic Meissner effect~\cite{2,3,4,5} (PME) originated
from a cooperative behavior of weak-links mediated orbital moments 
and found to be responsible for unusual aging effects~\cite{6} in high-$T_c$ granular 
superconductors (HTGS). Among others are also recently introduced thermophase~\cite{7,8} and 
piezophase~\cite{9} effects suggesting, respectively, a direct influence
of a thermal gradient and an applied stress on phase difference between the 
adjacent grains. Besides, using a model of random overdamped Josephson
junction arrays, two dual time-parity violating effects in HTGS have been 
predicted~\cite{10,11}. Namely, an appearance of magnetic field induced electric 
polarization along with the concomitant change of the junction capacitance 
(magnetoelectric effect~\cite{10}) and existence of electric field 
induced magnetization (converse magnetoelectric effect~\cite{11}) via a 
Dzyaloshinski-Moria type interaction mechanism.  

In this Letter we discuss a possibility of two other interesting effects expected 
to occur in a granular material under sufficient mechanical loading. 
Specifically, we predict the existence of stress induced paramagnetic moment in 
zero applied magnetic field (piezomagnetism) and its influence on a low-field 
magnetization (leading to a mechanically induced PME). 

The possibility to observe tangible piezoeffects in mechanically loaded grain
boundary Josephson junctions (GBJJs) is based on the following arguments. 
It is well known~\cite{12,13,14} that the grain boundaries (GBs) are the natural sources 
of weak links (or GBJJs) in granular superconductors. Under plastic deformation, GBs 
were found~\cite{15} to move rather rapidly via the movement of the grain boundary 
dislocations (GBDs) comprising these GBs. As a matter of fact, using the so-called
method of acoustic emission, the plastic flow of GBDs with the maximum rate of 
$v_0=1mm/s$ has been registered~\cite{16} in $YBCO$ ceramics at $T=77K$ under the 
external load of $\sigma =10^7N/m^2$. Using the above evidence, in Ref.9 a {\it piezophase} 
response of a single GBJJ (created by GBDs strain field $\epsilon _d$ acting as an 
insulating barrier of thickness $l$ and height $U$ in a $SIS$-type junction with the 
Josephson energy $J\propto e^{-l\sqrt{U}}$) to an externally applied mechanical stress was
considered. The resulting stress-strain and stress-current diagrams were found~\cite{9}  
to exhibit a quasi-periodic (Fraunhofer-like) behavior typical for Josephson junctions (JJs). 
To understand how piezoeffects can manifest themselves through GBJJs,
let us invoke an analogy with the so-called {\it thermophase effect} suggested originally 
by Guttman {\it et al}.~\cite{7} (as a quantum mechanical alternative for the conventional 
thermoelectric effect) to occur in a single JJ and later applied to HTGS~\cite{8}.
In essence, the thermophase effect assumes a direct coupling between an applied
temperature drop $\Delta T$ and the resulting phase difference $\Delta \phi$
through a JJ. When a rather small temperature gradient is applied to
a JJ, an entropy-carrying normal current $I_n=L_n\Delta T$ (where
$L_n$ is the thermoelectric coefficient) is generated through such a junction. 
To satisfy the constraint dictated by the Meissner effect, the resulting supercurrent 
$I_s=I_c\sin [\Delta \phi ]$ (with $I_c=2eJ/h$ being the Josephson critical current) 
develops a phase difference through a weak link. In other words, the temperature 
gradient stimulates a superconducting phase gradient which in turn drives the reverse
supercurrent. The normal current is locally canceled by a counterflow of 
supercurrent, so that the total current through the junction $I=I_n+I_s=0$. As a
result, supercurrent $I_c\sin [\Delta \phi ]=-I_n=-L_n\Delta T$ generates a nonzero 
phase difference
via a transient Seebeck thermoelectric field leading to the linear thermophase 
effect~\cite{7} $\Delta \phi =-\arcsin (L_{tp}\Delta T)\simeq -L_{tp}\Delta T$ 
with $L_{tp}=L_n/I_c(T)$.

By analogy, we can introduce a {\it piezophase effect} (as a quantum 
alternative for the conventional piezoelectric effect) through a JJ~\cite{9}.
Indeed, a linear conventional piezoelectric effect relates induced polarization $P_n$ 
to an applied strain $\epsilon$ as~\cite{17} $P_n=d_n\epsilon$, where $d_n$ is the 
piezoelectric coefficient. The corresponding normal piezocurrent density is $j_n=dP_n/dt=
d_n\dot {\epsilon}$ where $\dot {\epsilon}(\sigma )$ is a rate of plastic
deformation (under an applied stress $\sigma$) which depends on the number
of GBDs of density $\rho$ and a mean dislocation rate $v_d$ as follows~\cite{18} 
$\dot {\epsilon}(\sigma )=b\rho v_d(\sigma )$ (where $b$ is the absolute value
of the appropriate Burgers vector). In turn, $v_d\simeq 
v_0(\sigma /\sigma _m)$ with $\sigma _m$ being the so-called ultimate stress.
To meet the requirements imposed by the Meissner effect, in response to the induced normal 
piezocurrent, the corresponding Josephson supercurrent of density $j_s=dP_s/dt=
j_c\sin [\Delta \phi ]$ should emerge within the contact. Here $P_s=-2enb$
is the Cooper pair's induced polarization with 
$n=N/V$ the pair number density, and $j_c=2ebJ/\hbar V$ is the critical current density.
The neutrality conditions ($j_n+j_s=0$ and $P_n+P_s=const$) will lead then to the linear 
piezophase effect $\Delta \phi =-\arcsin [d_{pp}\dot {\epsilon}(\sigma )]
\simeq -d_{pp}\dot {\epsilon}(\sigma )$ (with $d_{pp}=d_n/j_c$ 
being the piezophase coefficient) and the concomitant change of the pair number density 
under an applied strain, viz., $\Delta n(\epsilon )=d_{pn}\epsilon$ with $d_{pn}=d_n/2eb$. 
Given the markedly different scales of stress induced changes in defect-free thin 
films~\cite{19} and weak-links-ridden ceramics~\cite{20}, it should be possible to 
experimentally register the suggested here piezophase effects.

To adequately describe magnetic properties of a granular superconductor, 
we employ a model of {\it random} three-dimensional (3D) overdamped Josephson junction 
array which is based on the well known tunneling Hamiltonian~\cite{21,22,23,24}
\begin{equation}
{\cal H}=\sum_{ij}^NJ(r_{ij})[1-\cos \phi_{ij}],
\end{equation}
where $\{i\}=\vec {r}_i$ is a 3D lattice vector, $N$ is the number of grains (or weak links),
$J(r_{ij})$ is the Josephson 
coupling energy with $\vec r_{ij}=\vec r_i-\vec r_j$ the separation
between the grains; the gauge invariant phase difference is defined as
$\phi _{ij}=\phi ^0_{ij}-A_{ij}$,
where $\phi ^0_{ij}=\phi _i-\phi _j$ with $\phi_i$ being the phase of the
superconducting order parameter, and 
$A_{ij}=\frac{2\pi}{\Phi_ 0}\int_i^j\vec A(\vec r)\cdot d{\vec l}$ is the frustration
parameter with $\vec A(\vec r)$ the electromagnetic vector potential which involves both 
external fields and possible self-field 
effects (see below); $\Phi_ 0=h/2e$ is the quantum of flux.

In the present paper, we consider a long-range interaction between grains~\cite{8,10,11,24} 
(with $J(r_{ij})=J$) and model the true short-range behavior 
of a HTGS sample through the randomness in the 
position of the superconducting grains in the array (see below). For simplicity, we shall  
ignore the role of Coulomb interaction effects assuming that the grain's charging 
energy $E_c\ll J$ (where $E_c=e^2/2C$, with $C$ the capacitance of the junction). 
As we shall see, this condition is reasonably satisfied for the effects under
discussion.

According to the above-discussed scenario, 
under an applied stress the superconducting phase 
difference will acquire an additional contribution 
$\delta \phi _{ij}(\sigma )=-B\vec \sigma\cdot \vec r_{ij}$, 
where $B=d_n\dot{\epsilon} _0/\sigma _mj_cb$ with $\dot{\epsilon} _0=b\rho v_0$ being 
the maximum deformation rate and the other parameters defined earlier.
If, in addition to the external loading, the network of 
superconducting grains is under the influence of an applied frustrating magnetic
field $\vec H$, the total phase difference through the contact reads (where $\vec R_{ij}=
(\vec r_i+\vec r_j)/2$)
\begin{equation}
\phi _{ij}(\vec H, \vec \sigma )=\phi ^0_{ij}+\frac{\pi}{\Phi _0}(\vec r_{ij}\wedge \vec R_{ij})
\cdot \vec H-B\vec \sigma \cdot \vec r_{ij}.
\end{equation}

It is well known~\cite{1,10,11,24} that the self-induced Josephson fields can in principle 
be quite pronounced for large-size junctions even in zero applied magnetic 
fields. So, to safely neglect the influence of these effects in a real
material, the corresponding Josephson penetration length $\lambda _J$ must be 
much larger than the junction (or grain) size. 
Specifically, this condition will be satisfied for short junctions with 
the size $d\ll \lambda _J$, where $\lambda _J=\sqrt{\Phi _0/4\pi \mu _0j_c 
\lambda _L}$ with $\lambda _L$ being 
the grain London penetration depth and $j_c$ its Josephson critical current 
density. In particular, since in HTGS 
$\lambda _L\simeq 150nm$, the above 
criterium will be rather well met for $d\simeq 1\mu m$ and $j_c\simeq 10^{4}A/m^2$
which are the typical parameters for HTGS ceramics~\cite{1}.
Likewise, to ensure the uniformity of the applied stress $\sigma$, we also 
assume that $d\ll \lambda _{\sigma}$, where $\lambda _{\sigma}$ is 
a characteristic length over which $\sigma$ is kept homogeneous. 

When the Josephson supercurrent $I_{ij}^s=I_c\sin \phi _{ij}$ 
circulates around a set of grains (that form a random area plaquette), it 
induces a random magnetic
moment $\vec \mu _s$ of the Josephson network~\cite{21}
\begin{equation}
\vec \mu _s\equiv -\frac{\partial {\cal H}}{\partial \vec H}=
\sum_{ij}I_{ij}^s(\vec r_{ij}\wedge \vec R_{ij}),
\end{equation}
which results in the stress induced net magnetization  
\begin{equation}
\vec M_s(\vec H,\vec \sigma)\equiv \frac{1}{V}<\vec \mu_s>=
\int\limits_{0}^{\infty }d\vec r_{ij}d\vec R_{ij}
f(\vec r_{ij}, \vec R_{ij}) \vec \mu _s,
\end{equation}
where $V$ is a sample's volume and $f$ the joint probability distribution 
function (see below). 
To capture the very essence of the superconducting piezomagnetic effect, in what follows we
assume for simplicity that an {\it unloaded sample} does not possess any spontaneous magnetization
at zero magnetic field (that is $M_s(0,0)=0$) and that its Meissner response to a small applied 
field $H$ is purely diamagnetic (that is $M_s(H,0)\simeq -H$). According to Eq.(2), this 
condition implies $\phi ^0_{ij}=2\pi m$ for the initial phase difference with $m=0,\pm 1, \pm 2,..$. 
Incidentally, this is also a requirement for current conservation at zero temperature~\cite{21}. 

In order to obtain an explicit expression for the piezomagnetization, we consider a 
site positional disorder that allows for small 
random radial displacements. Namely, the sites in a 3D cubic lattice are 
assumed to move from their equilibrium positions according to the normalized 
(separable) distribution function 
$f(\vec r_{ij}\vec R_{ij})\equiv f_{r}(\vec r_{ij})f_{R}(\vec R_{ij})$.
As usual~\cite{8,10,11}, it can be shown that the main qualitative results of this paper do 
not depend 
on the particular choice of the probability distribution function. 
For simplicity here we assume an exponential distribution law for the 
distance between grains, $f_r(\vec r)=f(x)f(y)f(z)$ with 
$f(x_j)=(1/d)e^{-x_j/d}$, and some
short range distribution for the 
dependence of the center-of-mass probability $f_R(\vec R)$ (around 
some constant value $D$). While the specific form of the latter distribution is not
important for the effects under discussion,  
it is worthwhile to mention that the former 
distribution function $f_r(\vec r)$ reflects a short-range character of the
Josephson coupling in granular superconductors where~\cite{25} 
$J(\vec r_{ij})=Je^{-\vec \kappa \cdot \vec r_{ij}}$. 
For isotropic arrangement of identical grains, with spacing 
$d$ between the centers of adjacent grains, we have 
$\vec \kappa =(\frac{1}{d},\frac{1}{d},\frac{1}{d})$ and thus $d$ is of the 
order of an average grain size. 

Taking the applied stress along the 
$x$-axis, $\vec \sigma=(\sigma,0,0)$, normally to the applied magnetic field
$\vec H=(0,0,H)$, we get finally
\begin{equation}
M_s(H,\sigma )=-M_0\frac{H_{tot}(H,\sigma )/H_0}{[1+H^2_{tot}(H,\sigma )/H^2_0]^2},
\end{equation}
for the induced transverse magnetization (along the $z$-axis), where
$H_{tot}(H,\sigma )=H-H^{*}(\sigma )$ is the total magnetic field with
$H^{*}(\sigma )=(\sigma /\sigma _0)H_0$  
being a stress-induced contribution. Here, $M_0=I_cSN/V$ with $S=\pi dD$ being 
a projected area around the Josephson contact,  
$H_0=\Phi _0/S$, and $\sigma _0=\sigma _m(j_c/j_d)(b/d)$ with $j_d=d_n\dot{\epsilon} _0$
and $\dot{\epsilon} _0=b\rho v_0$ being the maximum values of the dislocation current density 
and the plastic deformation rate, respectively.

Fig.1 presents the stress induced magnetization at different applied 
magnetic fields, calculated according to Eq.(5). As is seen, in practically zero magnetic 
field the piezomagnetization is purely paramagnetic (solid line), exhibiting a strong 
nonlinear behavior. With increasing the stress, it first increases reaching a maximum, 
and then rather rapidly dies away. Under the influence of small applied magnetic fields 
(dotted and dashed lines), the piezomagnetism turns diamagnetic (for low external stress) 
with its 
peak shifting toward higher loading. At the same time, Fig.2 shows changes of the initial 
stress-free diamagnetic magnetization (solid line) under an applied stress.
As we see, already relatively small values of an applied stress render a low
field Meissner phase strongly paramagnetic (dotted and dashed lines)
simultaneously shifting the peak toward higher magnetic fields. 
According to Eq.(5), the initially diamagnetic Meissner
effect turns paramagnetic as soon as the piezomagnetic contribution 
$H^{*}(\sigma )$ exceeds an applied magnetic field $H$. To see whether this 
can actually happen in a real material, let us estimate the typical values of the 
piezomagnetic field $H^{*}$. By definition, $H^{*}(\sigma )=
(\sigma /\sigma _m)(j_d/j_c)(d/b)H_0$ where $H_0=\Phi _0/S$ is a characteristic magnetic
field, and $\sigma _m$ is an ultimate stress field. 
Typically~\cite{3,4,5}, for HTGS ceramics $S\approx 10\mu m^2$, leading to $H_0\simeq 1G$. 
To estimate the needed value of the dislocation current density $j_d$, we turn to the
available experimental data. According to Ref.14, a rather strong polarization  
under compressive pressure $\sigma /\sigma _m \simeq 0.1$ was observed in $YBCO$ ceramic 
samples at $T=77K$ yielding $d_n=10^2C/m^2$ for the piezoelectric coefficient.
Usually~\cite{12,13,14,16,20}, for GBJJs $\dot{\epsilon} _0\simeq 10^{-2}s^{-1}$, and $b\simeq 10nm$
leading to $j_d=d_n\dot{\epsilon} _0\simeq 1A/m^2$ for the maximum dislocation current density.  
Using the typical values of the critical current density $j_c=10^4A/m^2$ and grain size 
$d\simeq 1\mu m$, we arrive at the following estimate of the piezomagnetic field 
$H^{*}\simeq 10^{-2}H_0$. Thus, the predicted stress induced paramagnetic Meissner effect 
(PME) should be observable for applied magnetic fields $H\simeq 10^{-2}H_0\simeq 0.01G$ 
which correspond to the region where the original PME was first registered~\cite{2,3,4,5,6}. 
In turn, the piezoelectric coefficient $d_n$ is related to an effective charge $Q$ in the GBJJ  
as~\cite{26} $d_n=(Q/S)(d/b)^2$. Given the above-obtained estimates, we get a 
reasonable value of $Q\simeq 10^{-13}C$ for the charge accumulated at a GBJJ.
It is interesting to notice that the above values of the aplied stress $\sigma $ 
and the resulting effective charge $Q$ correspond (via the so-called electroplastic 
effect~\cite{26}) to an equivalent applied electric field $E=b^2\sigma /Q\simeq 10^7V/m$
at which rather pronounced electric-field induced effects in HTGS were either observed 
(like an increase of the critical current in $YBCO$ ceramics~\cite{27}) or predicted to occur 
(like a converse magnetoelectric effect~\cite{11}).
 
In conclusion, let us briefly discuss the contribution of the so-called striction 
effects~\cite{23} (which usually accompany any stress related changes). According to Ref.28
the Josephson projected area $S$ was found to slightly decrease under pressure
thus leading to some increase of the characteristic field $H_0=\Phi _0/S$. In view of
Eq.(5), it means that a smaller compression  
stress will be needed to actually reverse the sign of the induced magnetization $M_s$. 
Furthermore, if an unloaded granular superconductor already exhibits the PME, due to the 
orbital currents induced spontaneous magnetization resulting from an initial phase difference 
$\phi ^0_{ij}=2\pi r$ in Eq.(2) with fractional $r$ (in particular, $r=1/2$ corresponds to 
the so-called~\cite{2,3,4,5,6} $\pi$-type state), then according to our predictions this 
effect will either be further enhanced by applying a compression (with $\sigma >0$)
or will disappear under a strong enough extension (with $\sigma <0$) able to compensate the 
pre-existing effect. Given a very distinctive nonlinear character of $M_s(H,\sigma )$ 
(see Figs.1 and 2), the above-estimated range of accessible parameters suggests quite an optimistic 
possibility to observe the predicted effects experimentally either in HTGS 
ceramics or in a specially prepared system of arrays of superconducting grains. Finally, it is 
worth noting that a rather strong nonlinear response of the transport properties in $HgBaCaCuO$ 
ceramics was observed~\cite{20} under compressive pressure with $\sigma /\sigma _m\simeq 0.8$. 
Specifically, the critical current at $\sigma =9 kbar$ was found to be three 
times higher its value at $\sigma =1.5 kbar$, clearly indicating a weak-links-mediated origin of 
the phenomenon (in the best defect-free thin films this ratio never exceeds a few percents~\cite{19}).

This work was done during my stay at ETH--Z\"urich and was funded by the Swiss National Science Foundation.
I thank Professor T.M. Rice for hospitality and stimulating discussions on the subject.

\begin{figure}
\epsfxsize=8.5cm
\centerline{\epsffile{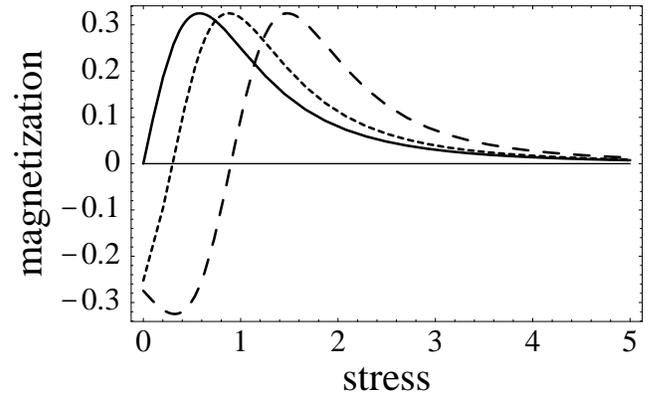} }
\caption{The induced magnetization $M_s/M_0$ as
a function of the reduced applied stress $\sigma /\sigma _0$,
according to Eq.(5) for different values of reduced applied magnetic field: 
$H/H_0=0.001$ (solid line), $H/H_0=0.01$ (dotted line), and $H/H_0=0.1$ 
(dashed line). }
\end{figure}

\begin{figure}
\epsfxsize=8.5cm
\centerline{\epsffile{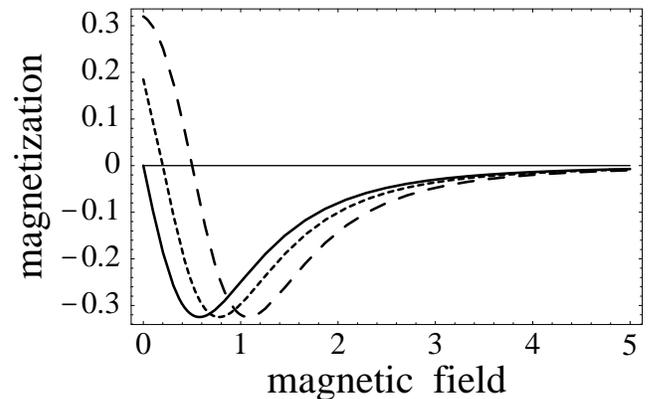} }
\caption{The magnetization $M_s/M_0$ as a function of the reduced applied 
magnetic field $H/H_0$, according to Eq.(5) for different values of reduced 
applied stress: $\sigma /\sigma _0=0$ (solid line), $\sigma /\sigma _0=0.01$ 
(dotted line), and $\sigma /\sigma _0=0.1$ (dashed line). }
\end{figure}

\begin{thebibliography}{99}
\bibitem[*]{byline} On leave from: Bogoliubov Laboratory of Theoretical Physics,
Joint Institute for Nuclear Research, 141980 Dubna, Russia
\bibitem{1} For recent reviews on the subject, see, e.g., {\em Macroscopic Quantum 
Phenomena and Coherence in Superconducting Networks}, ed. by C. Giovannella and M. Tinkham 
(World Scientific, Singapore, 1995) and {\em Superconductivity in Networks and 
Mesoscopic Structures}, ed. by C. Giovannella and C. Lambert  
(AIP Conference Proceedings $\#427$, 1998).
\bibitem{2} W. Braunisch {\it et al.}, Phys. Rev. Lett. {\bf 68}, 1908 (1992);
F.V. Kusmartsev, Phys. Rev. Lett. {\bf 69}, 2268 (1992).
\bibitem{3}  D. Khomski, J. Low Temp. Phys. {\bf 95}, 205 (1994).
\bibitem{4} J.R. Kirtley {\it et al.}, J. Phys.: Cond. Mat.
{\bf 10}, L97 (1998).
\bibitem{5}  M. Sigrist and T.M. Rice, Rev. Mod. Phys. {\bf 67}, 503 (1995).
\bibitem{6} E.L. Papadopoulou, P. Nordbad, and P. Svedlindh, Phys. Rev. Lett. {\bf 82}, 173 (1999).
\bibitem{7}  G.D. Guttman {\it et al.}, Phys. Rev. B 
{\bf 55}, 3849, 12691 (1997); ibid {\bf 57}, 2717 (1998).
\bibitem{8}  S. Sergeenkov, JETP Lett. {\bf 67}, 680 (1998).
\bibitem{9}  S. Sergeenkov, J. Phys.: Cond. Mat. {\bf 10}, L265 (1998).
\bibitem{10} S. Sergeenkov, J. Phys. I France {\bf 7}, 1175 (1997).
\bibitem{11}  S.A. Sergeenkov and J.V. Jos\'e, Europhys. Lett. 
{\bf 43}, 469 (1998).
\bibitem{12} M. Chisholm and S.J. Pennycook, Nature (London) {\bf 351},
47 (1991).
\bibitem{13} E. Sarnelli, P. Chaudhari, and J. Lacey, Appl. Phys. Lett.
{\bf 62}, 777 (1993).
\bibitem{14} T.J. Kim, E. Mohler, and W. Grill, J. Alloys Compd.
{\bf 211/212}, 318 (1994).
\bibitem{15} Y. Zhu {\it et al.}, Appl. Phys. Lett. {\bf 54}, 374 (1989).
\bibitem{16}  V.N. Kovalyova {\it et al.}, Sov. J. Low Temp. Phys. {\bf 17}, 46 (1991).
\bibitem{17} L.D. Landau and E.M. Lifshitz, {\em Electrodynamics
of Continuous Media} (Pergamon Press, New York, 1984).
\bibitem{18}  A.H. Cottrell, {\it Dislocations and  Flow in Crystals}
(Clarendon Press, Oxford, 1953).
\bibitem{19} G.L. Belenky {\it et al.}, Phys. Rev. B {\bf 44}, 10117 (1991).
\bibitem{20} A.I. D'yachenko {\it et al.}, Physica {\bf 251C}, 207 (1995).
\bibitem{21} C. Ebner and D. Stroud, Phys. Rev. B {\bf 31}, 165 (1985).
\bibitem{22} J. Choi and J.V. Jos\'e, Phys. Rev. Lett. {\bf 62}, 320 (1989).
\bibitem{23} S. Sergeenkov and M. Ausloos, Phys. Rev. B {\bf 48}, 604 (1993). 
\bibitem{24} G. Blatter {\it et al.}, Rev. Mod. Phys. {\bf 66}, 1125 (1994);
V.M. Vinokur {\it et al.}, Sov. Phys. JETP {\bf 66}, 198 (1987); H.R. Harbaugh and
D. Stroud, Phys. Rev. B {\bf 58}, R14759 (1998); H.R. Shea and M. Tinkham, Phys. 
Rev. Lett. {\bf 79}, 2324 (1997).
\bibitem{25} B. M\"uhlschlegel and D.L. Mills, Phys. Rev. B {\bf 29}, 159 (1984).
\bibitem{26} Yu. Ossipian {\it et al.}, Adv. Phys. {\bf 35}, 115 (1986).
\bibitem{27} T.S. Orlova and B.I. Smirnov, Supercond. Sci. Technol. {\bf 7}, 899
(1994).
\bibitem{28} J. Stankowski {\it et al.}, Physica {\bf 160C}, 170 (1989).
\end{thebibliography}
\end{document}